\documentclass[preprint,12pt]{elsarticle}

\usepackage[utf8]{inputenc} 
\usepackage[T1]{fontenc}    
\usepackage{hyperref}       
\usepackage{url}            
\usepackage{booktabs}       
\usepackage{amsfonts}       
\usepackage{nicefrac}       
\usepackage{microtype}      
\usepackage{xcolor}         
\usepackage{subcaption}     
\usepackage{graphicx}       
\definecolor{ocean}{RGB}{30,150,180}

\usepackage{comment}

\usepackage{amsmath, amssymb}
\usepackage{enumitem}
\usepackage{wrapfig,lipsum,booktabs}
\usepackage{float}
\usepackage{multirow}
\usepackage{longtable}
\usepackage{booktabs}

\usepackage{lineno}

\journal{Neural Networks}

\begin{document}

\begin{frontmatter}



\title{Image-based Data Representations of Time Series: A Comparative Analysis in EEG Artifact Detection}

\author[osnabrück]{Aaron Maiwald\fnref{co}}
\author[osnabrück]{Leon Ackermann\fnref{co}}
            
\author[osnabrück]{Maximilian Kalcher}
            
\author[stanford]{Daniel J. Wu\corref{corresponding}}
\ead{danjwu@cs.stanford.edu}
\cortext[corresponding]{Corresponding Author.}
\fntext[co]{These authors contributed equally to this work.}
\affiliation[osnabrück]{organization={University of Osnabrück Department of Cognitive Science},
addressline={Wachsbleiche 27},
city={Osnabrück},
postcode={49090},
state={Lower Saxony},
country={Germany}}
\affiliation[stanford]{organization={Stanford University Department of Computer Science},
addressline={353 Jane Stanford Way},
city={Stanford},
postcode={94305},
state={CA},
country={USA}}



\begin{abstract}
Alternative data representations are powerful tools that augment the performance of downstream models. However, there is an abundance of such representations within the machine learning toolbox, and the field lacks a comparative understanding of the suitability of each representation method.

In this paper, we propose artifact detection and classification within EEG data as a testbed for profiling image-based data representations of time series data. We then evaluate eleven popular deep learning architectures on each of six commonly-used representation methods.

We find that, while the choice of representation entails a choice within the tradeoff between bias and variance, certain representations are practically more effective in highlighting features which increase the signal-to-noise ratio of the data. We present our results on EEG data, and open-source our testing framework to enable future comparative analyses in this vein.

\end{abstract}




\end{frontmatter}


\pagebreak
\section{Introduction}

Data transformations are essential tools for machine learning researchers and practitioners. The transformation of raw data into alternative representations often unlocks downstream learning. These transformations are not just academic exercises; they are essential tools, with proven effectiveness in enhancing interpretability, efficiency, and model performance across a wide variety of disciplines.

\begin{figure}[H]
    \centering
    \begin{subfigure}{0.25\textwidth}
        \centering
        \includegraphics[width=\textwidth]{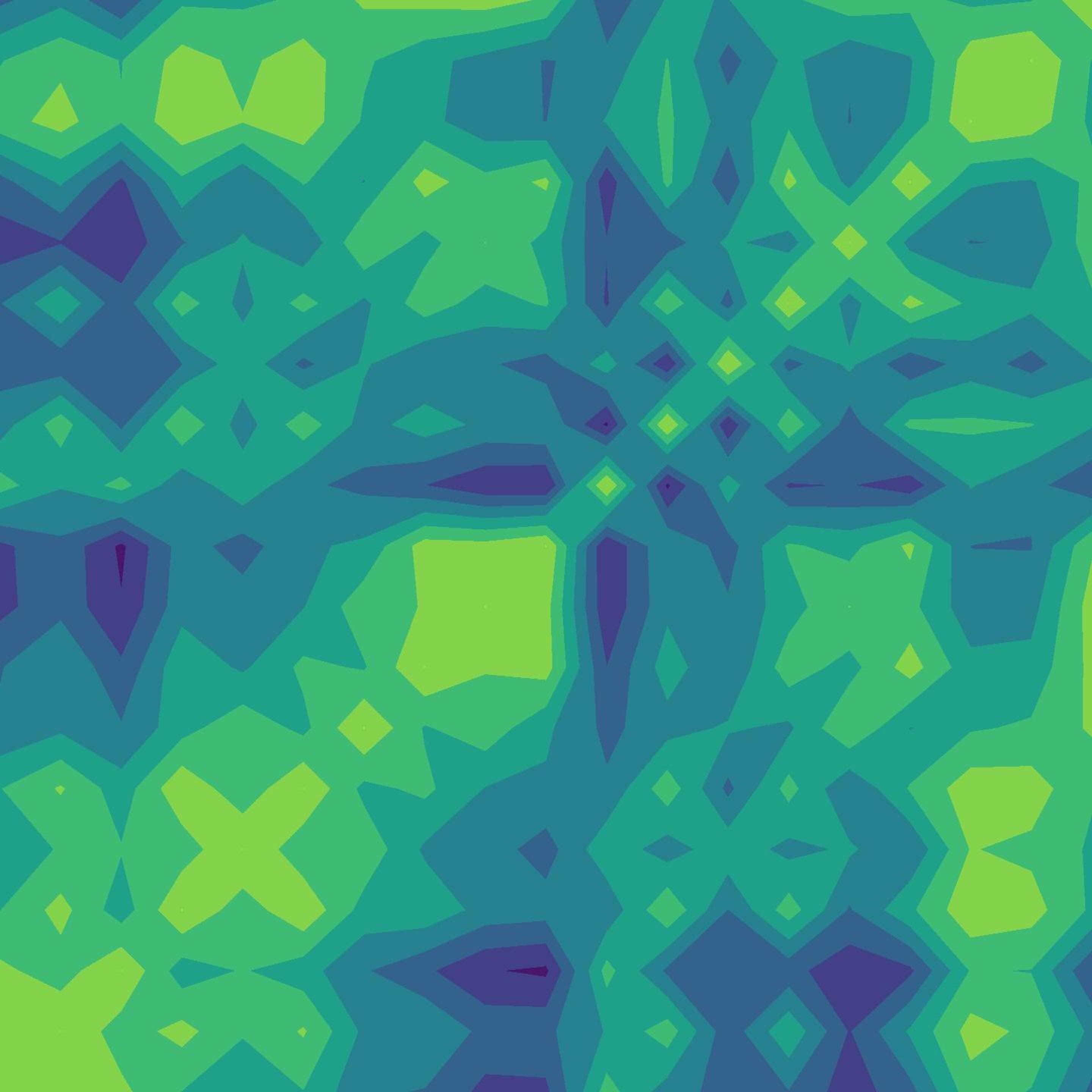}
        \caption{Correlation Matrix}
        \label{fig:corr}
    \end{subfigure}
    \begin{subfigure}{0.25\textwidth}
        \centering
        \includegraphics[width=\textwidth]{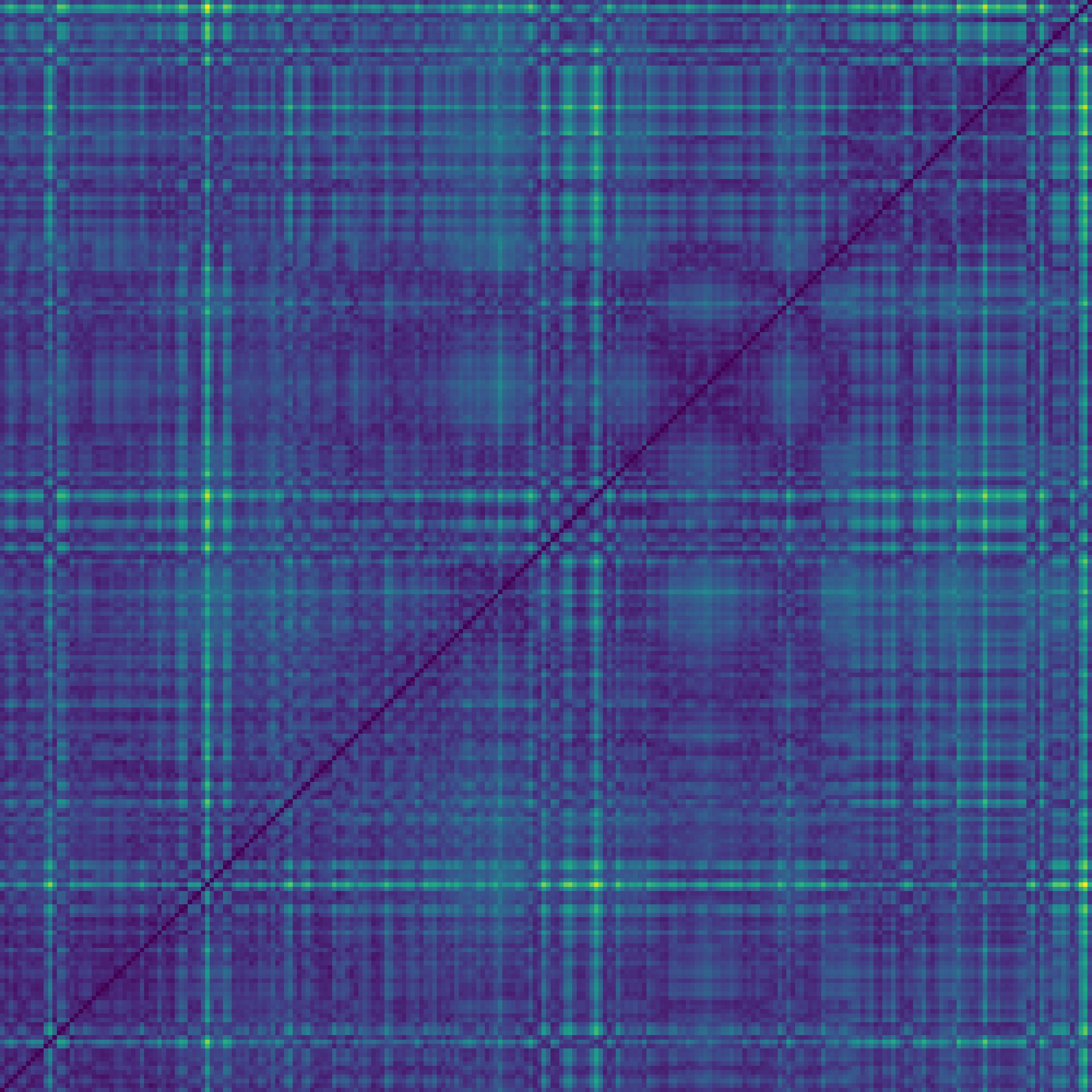}
        \caption{Recurrence Plot}
        \label{fig:recurrence}
    \end{subfigure}
    \begin{subfigure}{0.25\textwidth}
        \centering
        \includegraphics[width=\textwidth]{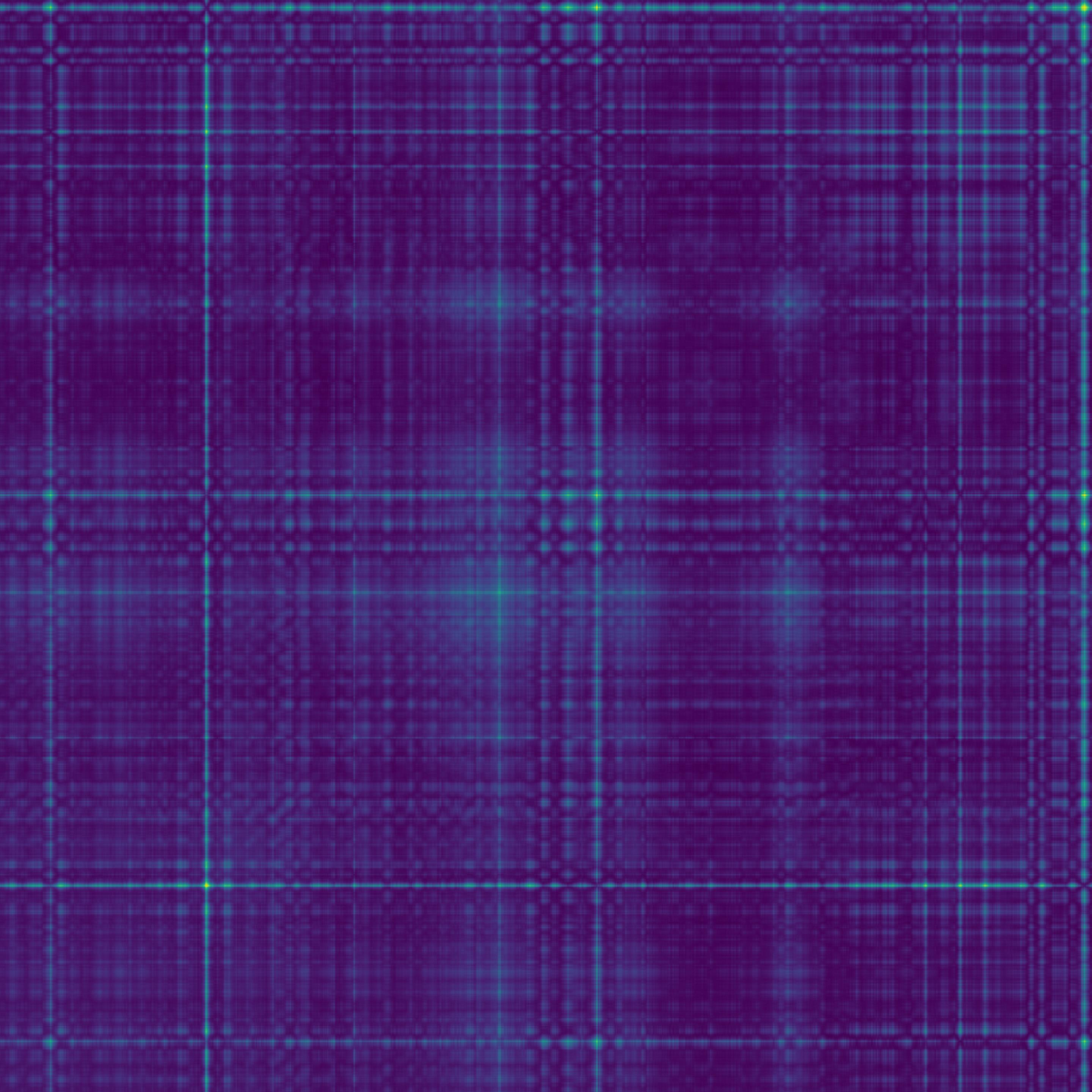}
        \caption{Gramian Angular Field}
        \label{fig:gramian}
    \end{subfigure}
    \hfill
    \begin{subfigure}{0.25\textwidth}
        \centering
        \includegraphics[width=\textwidth]{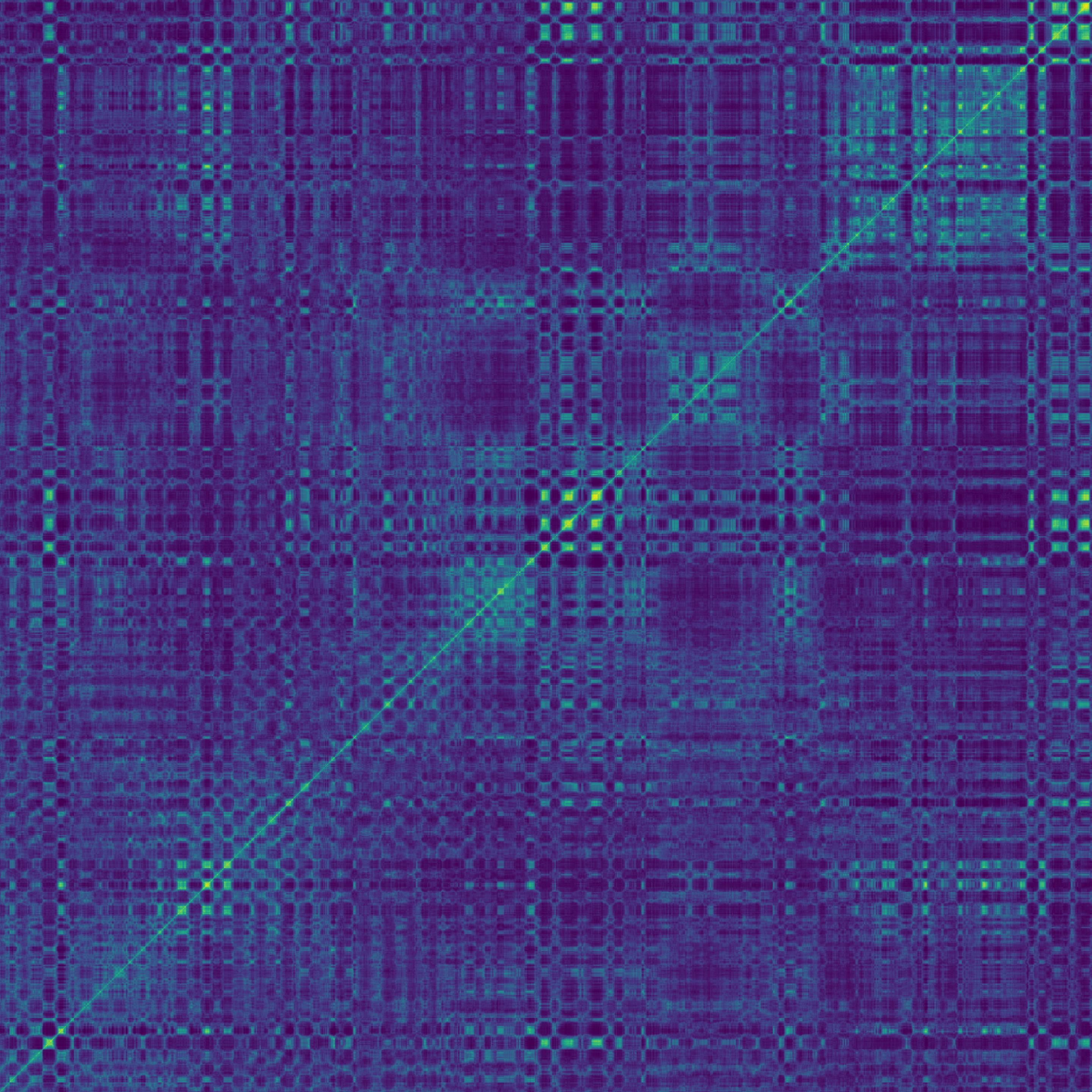}
        \caption{Markov Transition}
        \label{fig:markov}
    \end{subfigure}
    \begin{subfigure}{0.25\textwidth}
        \centering
        \includegraphics[width=\textwidth]{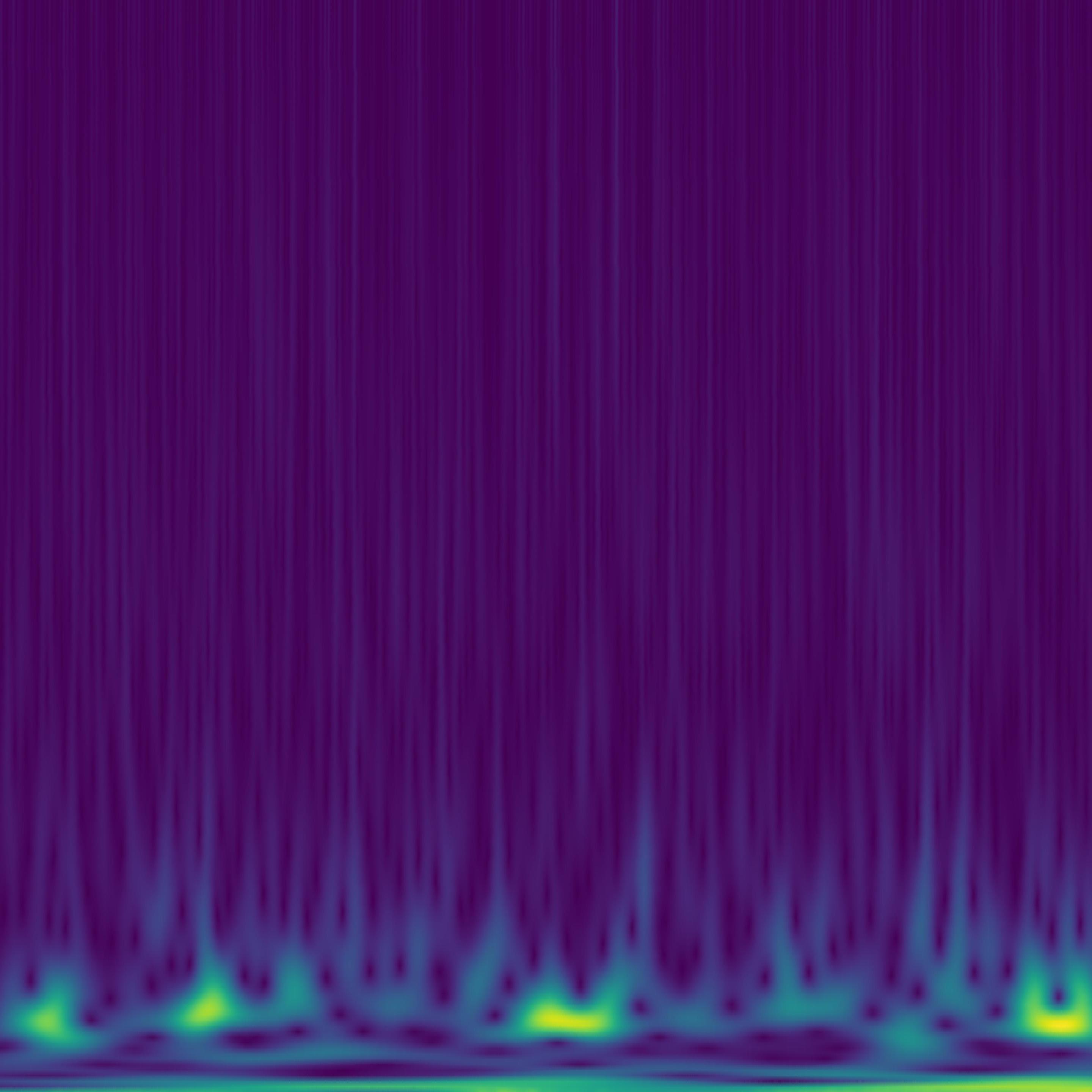}
        \caption{Continuous Wavelet}
        \label{fig:cwt}
    \end{subfigure}
    \begin{subfigure}{0.25\textwidth}
        \centering
        \includegraphics[width=\textwidth]{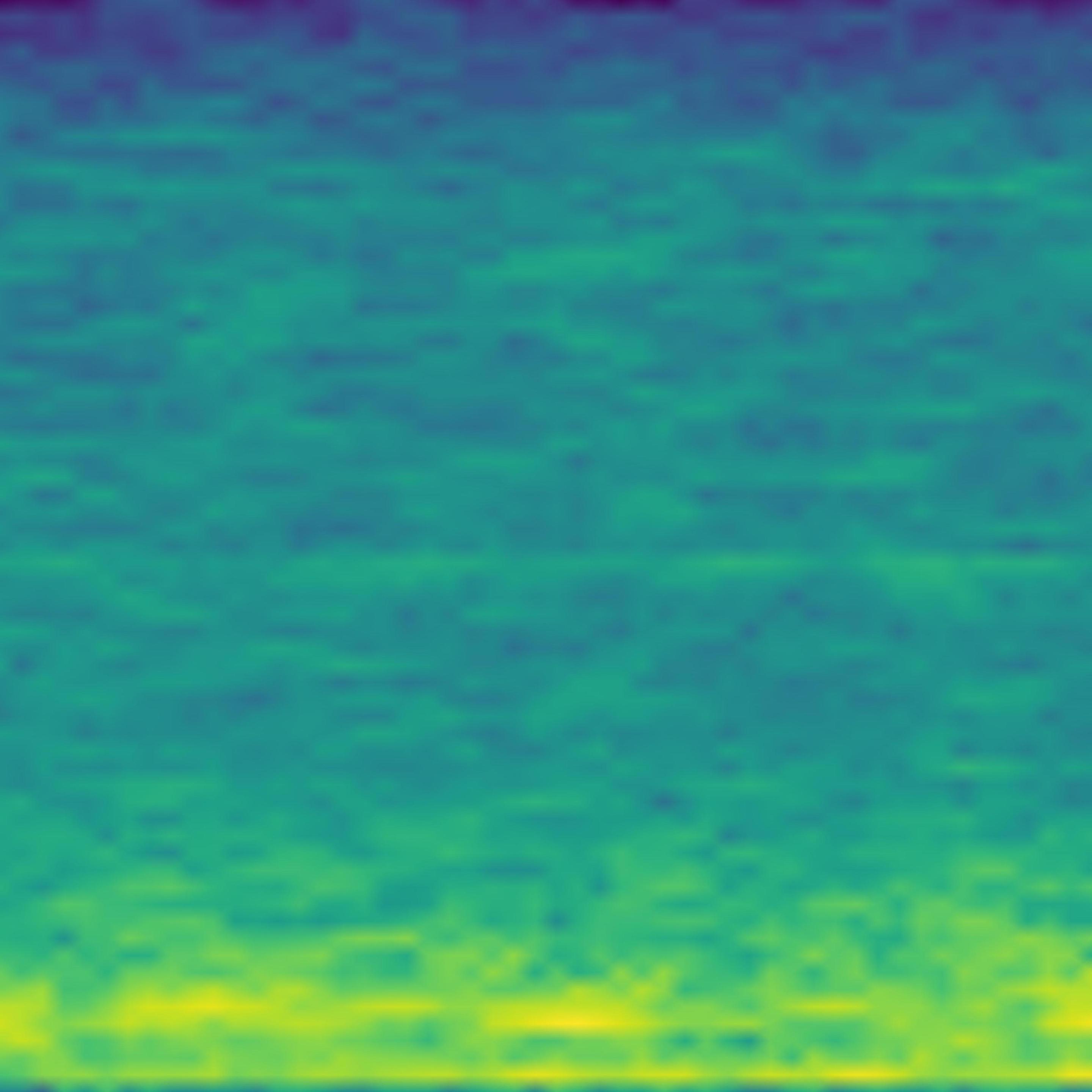}
        \caption{Spectrogram}
        \label{fig:spectrogram}
    \end{subfigure}

    \caption{Exemplar image representations of time series EEG data.}
    \label{fig:transforms}
\end{figure}

There is a wide panoply of such transformations within the research scientists' toolbox, and a comparative study of such transformations is useful for enabling efficent and effective future research.

As such, our work makes three primary contributions:
\begin{itemize}
    \item We introduce artifact detection in EEG data as a useful toy classification problem for assessing alternative data representation methods.
    \item We profile the characteristics and performance of six time series data representation methods on our toy problem, across a wide range of deep learning model architectures.
    \item We opensource our testing framework and testbed to facilitate future investigations into data representation methods.
\end{itemize}

\begin{wrapfigure}{r}{0.50\textwidth}
    \centering
    \begin{subfigure}{0.49\textwidth}
        \centering
        \includegraphics[width=\textwidth]{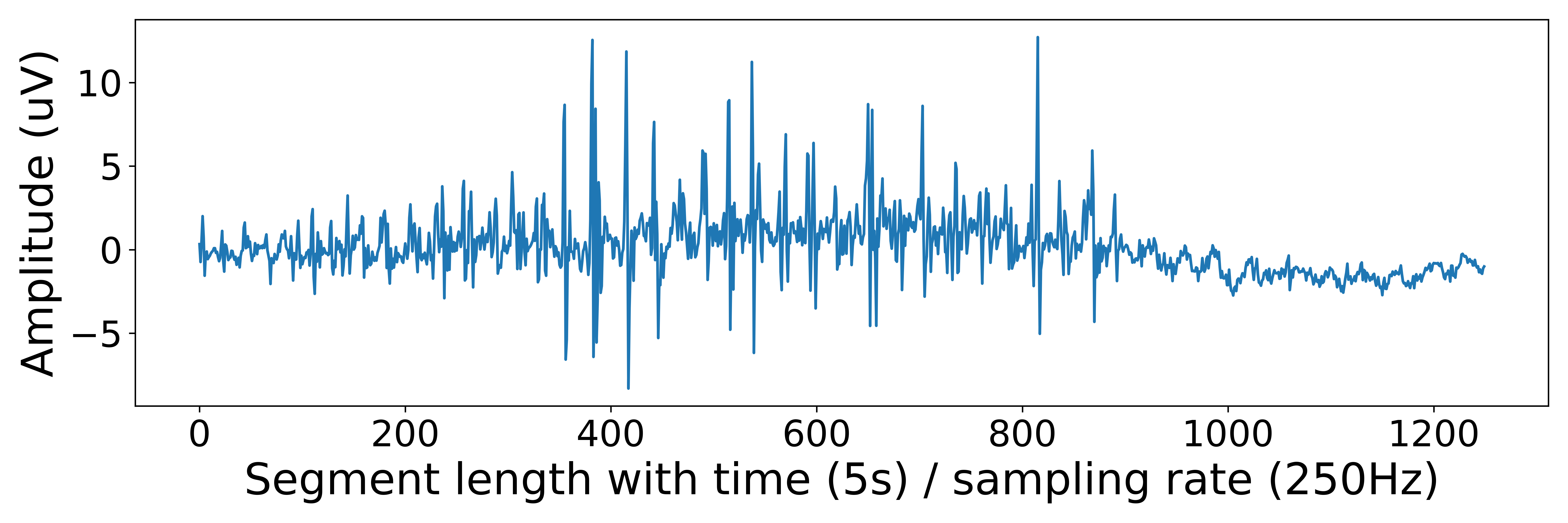}
        \caption{Chewing artifact}
    \end{subfigure}
    \begin{subfigure}{0.49\textwidth}
        \centering
        \includegraphics[width=\textwidth]{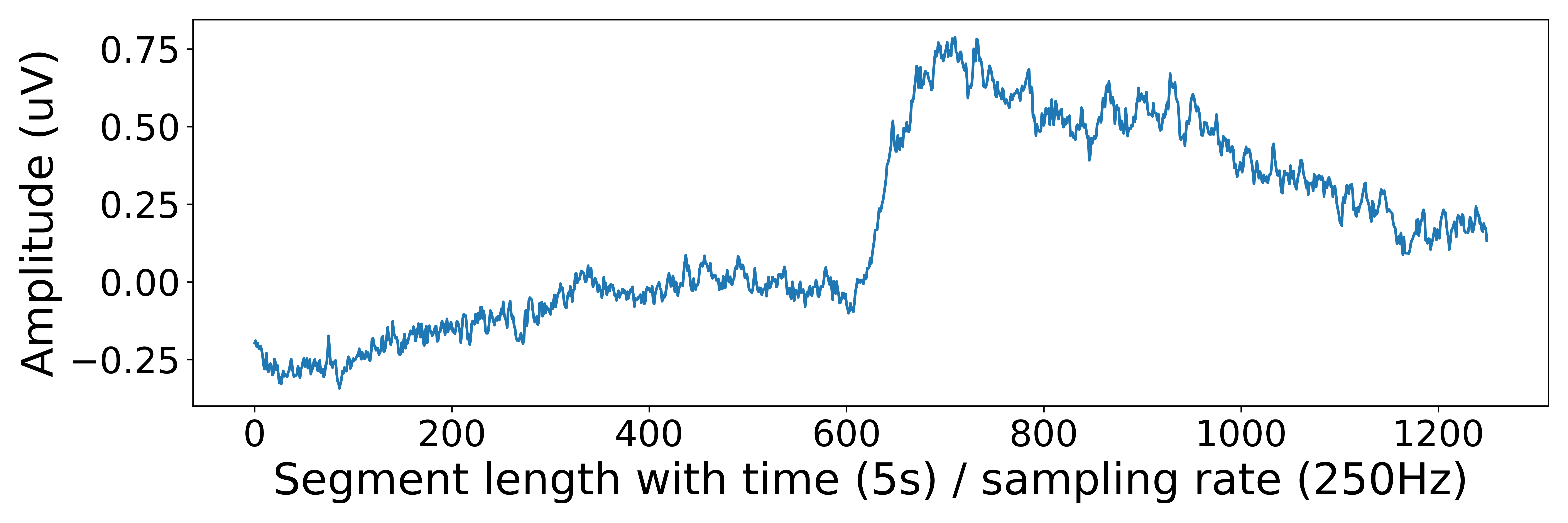}
        \caption{Electrode artifact}
    \end{subfigure}
    \begin{subfigure}{0.49\textwidth}
        \centering
        \includegraphics[width=\textwidth]{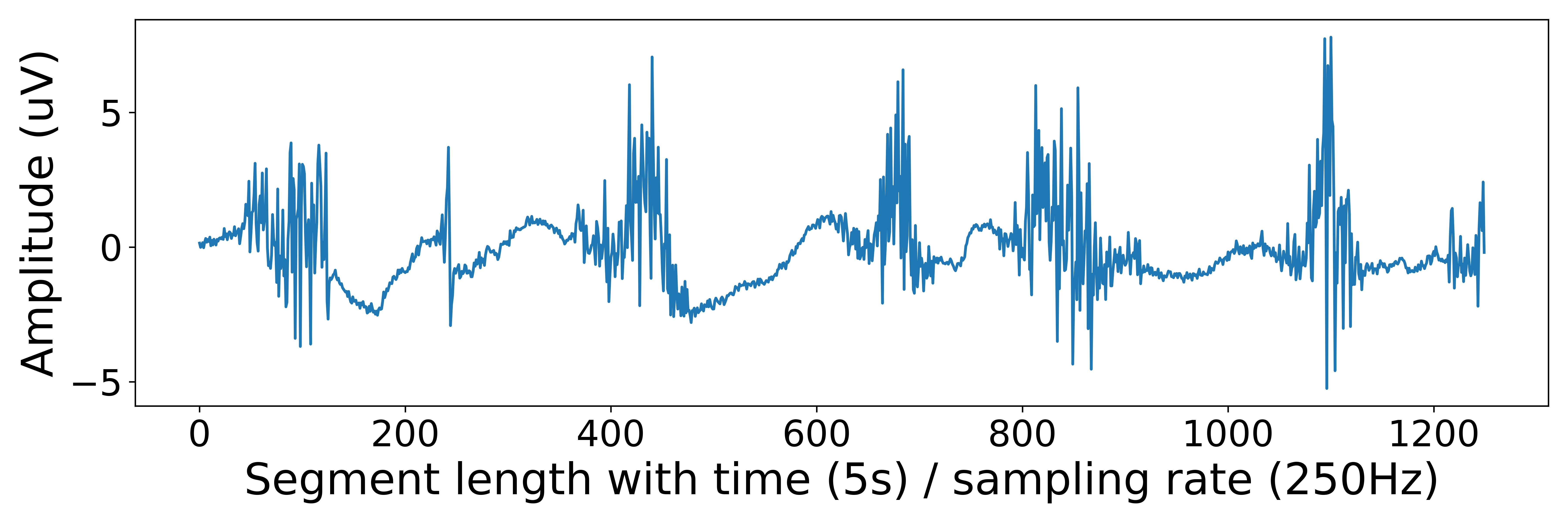}
        \caption{Eye movement artifact}
    \end{subfigure}
    \begin{subfigure}{0.49\textwidth}
        \centering
        \includegraphics[width=\textwidth]{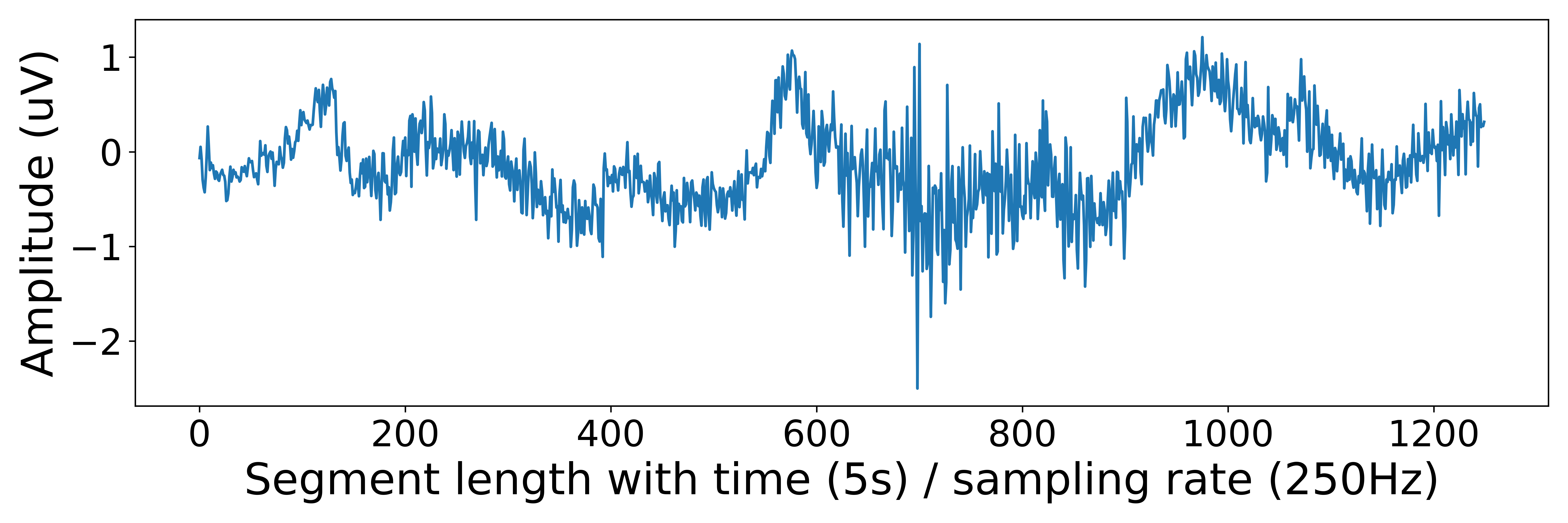}
        \caption{Muscle artifact}
    \end{subfigure}
    \begin{subfigure}{0.49\textwidth}
        \centering
        \includegraphics[width=\textwidth]{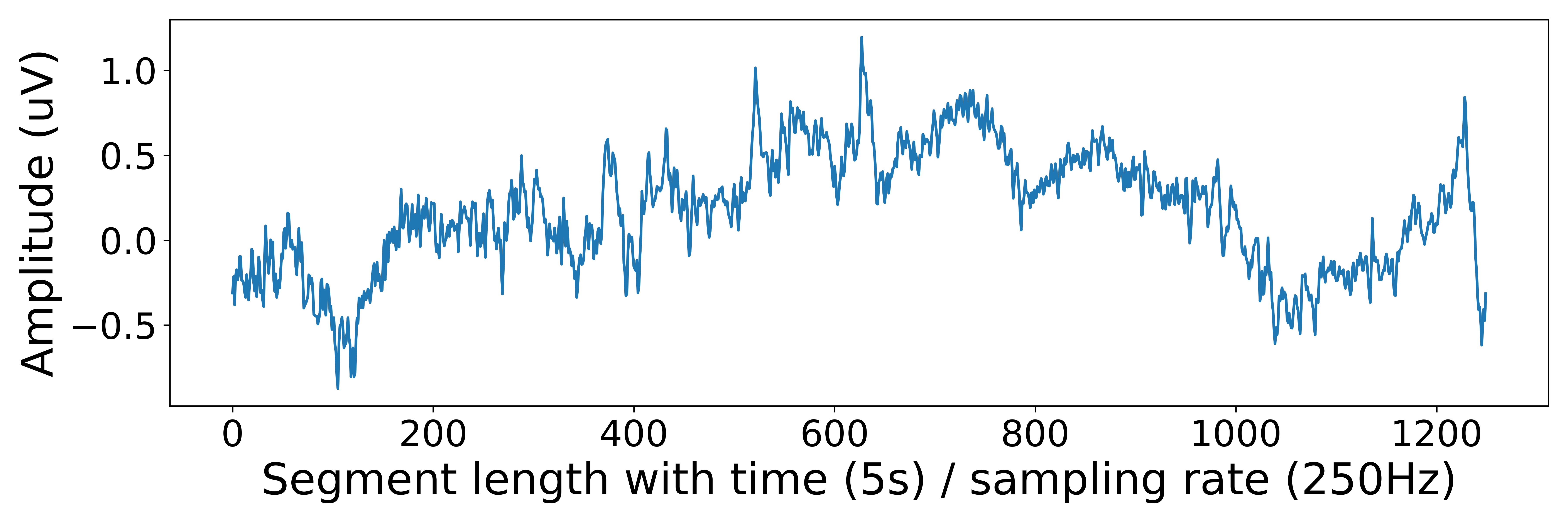}
        \caption{Shivering artifact}
    \end{subfigure}
    \caption{Different artifacts in EEG data.}
    \label{fig:artifacts}
\end{wrapfigure}

In Section \ref{sec:related}, we introduce alternative data representations in machine learning, and a toy problem to analyze time series data representations. Section \ref{sec:methods} presents our dataset and six data representation methods to be examined, while Section \ref{sec:experiments} describes our experimental pipeline. Section \ref{sec:results} presents the results of our experiments, which are discussed in Section \ref{sec:discussion}. Section \ref{sec:conclusion} concludes.

\section{Related Work}
\label{sec:related}

In this section, we examine the breadth of data transformation methods used in machine learning, before turning our attention to the specific domain in which we are interested: image representations of time series data. Finally, we introduce a toy problem, artifact detection within EEG data, which forms a testbed for comparative analysis of these image representations.

\subsection{Alternative Data Representations}

Alternative representations of raw data are produced by various data transformation pipelines. Data augmentation is a well-known application of these representations; augumented datasets bolster accuracy and mitigate overfitting \citep{shorten2019survey}. 

Alternative representations may also make latent information in raw data amenable to downstream machine learning. In early computer vision research, filter-based feature extraction was essential to tree-based model understanding of image data \citep{he2012guided}. Similarly, early work in speech processing represented voice data as mel-frequency cepstra to enable auditory information processing \citep{davis1980comparison}. More recently, methods like node2vec have been developed to convert graph-based data into vectors for consumption by language models \citep{grover2016node2vec}. Similarly, word2vec, and other word embedding engines, represent words as vectors for neural language models \citep{mikolov2013efficient}.

\subsection{Image Representations of Time Series Data}

Many signal processing methods, such as short Fourier transforms, mel-spectrograms, cochleagrams, and continuous wavelet transforms, have been used to convert sequential time series data into an image for visualization and analysis.

Individual image representations of time series data have been used with success in various domains, including speech \citep{arias2021multi}, sensor data \citep{yang2019multivariate}, and EEG data \citep{Bahador2020Correlation}.

However, in this diverse landscape of image representations for sequence data, there is no clear winner. While many studies have shown the utility of transforming time series data into images, to our best knowledge, the field lacks comprehensive comparative analyses between these time series representation methods.

\subsection{Artifact Detection in EEG data}

In order to make strides towards such a comparative analysis, it is useful to introduce a toy task: artifact detection within EEG data.

Electroencephalography (EEG) is a valuable non-invasive method for recording the electrical activity of the brain. EEG offers high temporal resolution, capturing brain activity on the order of milliseconds \citep{michel_eeg_2019}. However, EEG recordings often suffer from noise artifacts, irrelevant signals originating from various sources, including eye movements, muscle activity, and external interference \citep{jiang_removal_2019}. Detecting and removing these artifacts is crucial for accurate EEG data analysis. 

EEG artifact detection serves as a useful toy problem for studying data representation techniques \citep{tiwary2021time}.

Firstly, EEG artifact detection shares common characteristics with real-world problems involving time-series data \citep{fu2011review}. Much like financial data, EEG data is periodic, but noisy. Much like sensor data, EEG data involves multichannel measurements at high temporal resolution. And much like speech recognition data, EEG data contains information rich across the frequency spectrum. Thus, we consider EEG data to be broadly representative of many time series classification tasks.

Secondly, EEG artifact detection is a well-explored research area in neuroscience. Researchers have successfully applied machine learning techniques, including deep learning models pretrained on unrelated datasets like ImageNet, to enhance EEG data analysis \cite{sadiq2022exploiting}. Convolutional neural networks (CNNs) have proven effective in transforming EEG data into visual representations, improving artifact detection \citep{xu2019deep, bahador_automatic_2020}. This toy problem is still challenging, but tractable, making it useful for such a comparative analysis of data representation methods.

\section{Methods}
\label{sec:methods}
\subsection{Dataset}

We train and evaluate our classifier on the TUH EEG Artifact Corpus (TUAR) from the Temple University Hospital of Philadelphia (TUH), a subset of the TUH EEG Corpus \cite{hamid_temple_2020}. The TUAR contains normal EEG signals, and EEG signals are affected by five types of artifacts: chewing events, eye movements, muscular artifacts, shivering events, and instrumental artifacts (such as electrode pop, electrostatic artifacts, or lead artifacts). These artifacts are subject to significant data imbalance; eye movement and muscle artifacts account for 45.7\% and 35.9\%, respectively. Electrode pops make up 15.9\%, while chewing and shivering are the rarest artifacts with 2.2\% and 0.1\%, respectively (see Figure \ref{fig:artifacts}). The majority of the data however consists of normal EEG signals. In total, the dataset comprises 259 EEG sessions collected from 213 patients between 10 and 90 years old, over a span of 13 years. The EEGs were recorded with a sampling frequency of 250 Hz and 16-bit resolution.

\subsection{Preprocessing}

To preprocess the data, we segment the continuous EEG recordings into overlapping windows of EEG data. We chose a window length of 5 seconds based on previous research \citep{peh_transformer_2022}. Since not all recordings had the same frequency, we downsampled them to a frequency of $128$Hz. We included all

\begin{figure}[!t]
  \centering
  \begin{subfigure}[t]{\textwidth}
    \centering
    \includegraphics[width=\textwidth]{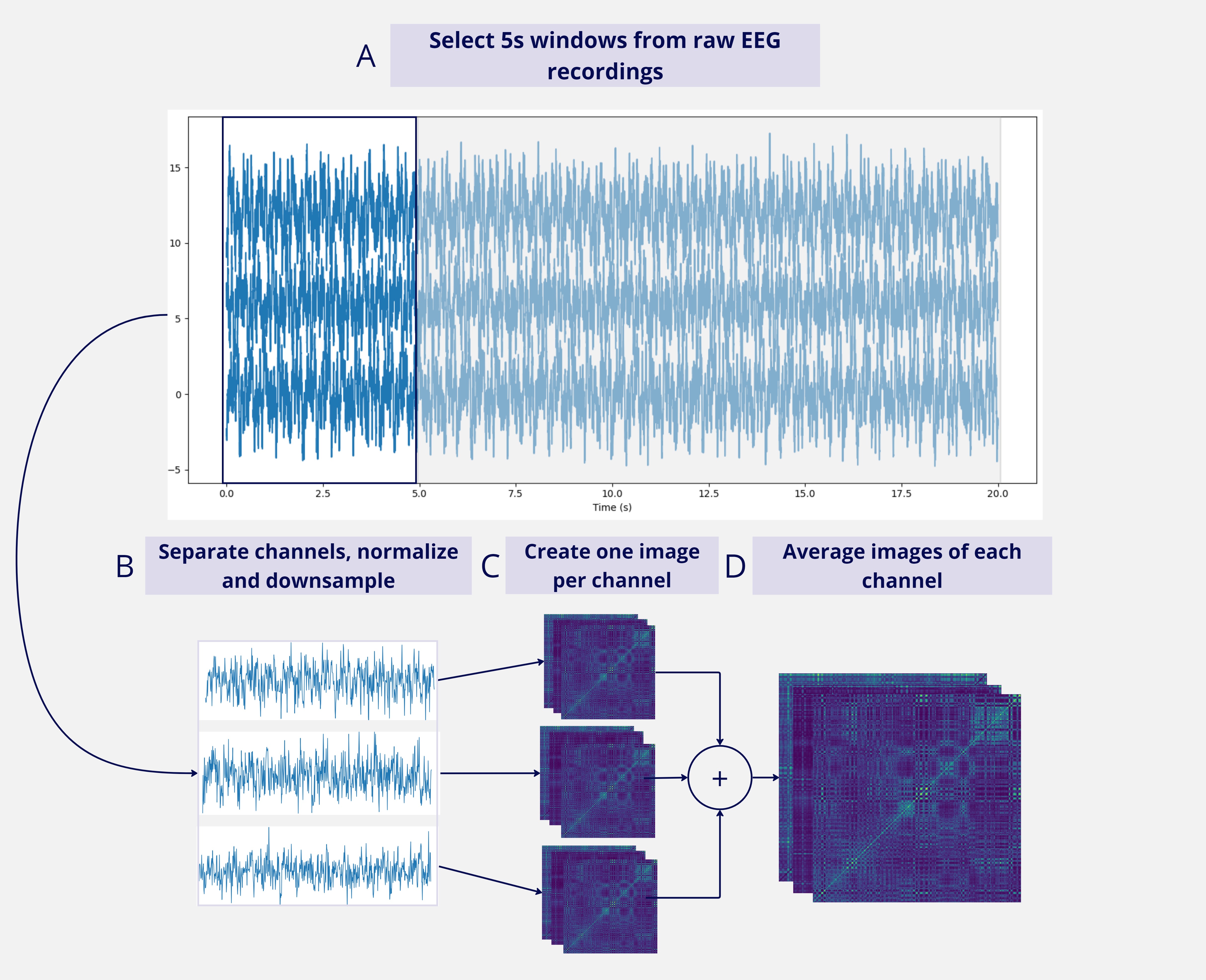}
  \end{subfigure}
\caption{Raw EEG data undergoes four preprocessing steps: (a) partitioning into 5s windows, (b) channel separation, normalization, and downsampling to 128 Hz, (c) creating visual representations for each channel, and (d) averaging them into a single image representing all channels (except for correlation matrices, which ensemble channels by default). For the purpose of illustration, we depict only 3 channels and 3 representation techniques.}
\label{fig:preprocessing}
\end{figure}

and only the channels that were present in all of the data, which amounted to 19 of over 50 channels. In line with common deep learning practice, we normalized the data with a $z$-transform (see Figure \ref{fig:preprocessing}).

\subsection{Time Series Data Representations}

We selected a number of data representation methods which are often used in the literature \citep{tiwary2021time, ismail2019deep, nakano2019effect}. Below, we introduce and define each method.

\subsubsection{Correlation Matrices}

Correlation matrices capture cross-channel similarities in multi-channel time series. We note that it is the only representation we examine that distills all channels into a single representation.

Formally, the correlation matrix $C$ for $n$ channels is defined as
\begin{equation}
C=
\begin{bmatrix}
\begin{array}{ccc}
\rho\left(\mathrm{S}_1, \mathrm{~S}_1\right) & \cdots & \rho\left(\mathrm{S}_1, \mathrm{~S}_{\mathrm{n}}\right) \\
\vdots & \ddots & \vdots \\
\rho\left(\mathrm{S}_{\mathrm{n}}, \mathrm{S}_1\right) & \cdots & \rho\left(\mathrm{S}_n, \mathrm{~S}_{\mathrm{n}}\right)
\end{array}
\end{bmatrix}
\end{equation}

where $\rho\left(\mathrm{S}_{\mathrm{i}}, \mathrm{S}_{\mathrm{j}}\right)$ is the correlation between channel $S_i$ and $S_j$.

The correlation matrix $C$ is visualized as a contour plot (see Figure \ref{fig:corr}). Correlation matrices have been successfully used to capture spatial dependencies and relationships between different electrodes \citep{Bahador2020Correlation}. 

\subsubsection{Recurrence Plots}
Recurrence plots are simple method to identify recurrent sequences in time series data \citep{nakano2019effect}.
Given a time series $X = \left(x_1, \ldots, x_N\right)$, the binarized recurrence matrix $R$ is given by

\begin{equation}
    R = 
    \begin{bmatrix}
    \mathbf{1}(|x_{1} - x_{N}| > \epsilon) & \dots & \mathbf{1}(|x_{1} - x_{1} |  > \epsilon)\\
    \vdots & \ddots & \vdots \\
    \mathbf{1}(|x_{N} - x_{N}|  > \epsilon)& \dots & \mathbf{1}(|x_{N} - x_{1}|  > \epsilon)\\
    \end{bmatrix} \\
\end{equation}

where $\epsilon$ is some binarization threshold value, and $\mathbf{1}(x)$ is the Heaviside step function, which takes the value of $1$ when the condition $x$ is true, and $0$ otherwise.

Thus, the recurrence plot $R$ is a binary matrix where a value of 0 at position $(i, j)$ indicates that data points $x_i$ and $x_j$ are considered recurrent (see Figure \ref{fig:recurrence}). These plots are commonly used \citep{ouyang2008using, bahari2013eeg} for EEG analysis, as normal EEG signals are consistently periodic, while abnormalities are aperiodic.

\subsubsection{Gramian Angular Summation Fields}

Gramian Angular Summation Fields (GASF) encode time series data into matrices that preserve temporal correlation information \citep{wang2015encoding}. In this representation, each data point $x_t \in \{x_1, x_2, ..., x_N\}$ at time $t$ is mapped to a polar coordinate with a mapping function $\phi$, given by:
\begin{equation}
    \phi(x_t) = \left(\frac{t}{N}, \arccos(x_t)\right)
\end{equation}

Then, the GASF $G$ is composed of the pairwise inner products of the angular representations of each data point:
\begin{equation}
    G = 
    \begin{bmatrix}
    \cos(\phi(x_1) + \phi(x_1)) & \dots & \cos(\phi(x_1) + \phi(x_N)) \\
    \vdots & \ddots & \vdots \\
    \cos(\phi(x_N) + \phi(x_1)) & \dots & \cos(\phi(x_N) + \phi(x_N)) \\
    \end{bmatrix} \\
\end{equation}

The resulting GASF $G$ is a 2D representation of the time series that captures its temporal correlations (see Figure \ref{fig:gramian}). These GASFs have proven to be successful in teasing out temporal trends in EEG data \citep{islam2019densenet, thanaraj2020implementation}.

\subsubsection{Markov Transition Fields}

Markov Transition Fields (MTF) are a matrix of transition probabilities between states of a time series \cite{wang2015spatially}. That is, given a time series $X = \{x_1, x_2, ..., x_N\}$, we divide the data into $Q$ quantiles, and then assign each $x_i$ to its corresponding bin $q \in [1, Q]$. Then, let us denote the bin containing $x_i$ to be $q_i$.

The Markov Transition Field $M$ is an $N\times N$ time-aware transition matrix, wherein each element $M_{ij}$ represents the transition probability from the quantiles $q_i \rightarrow q_j$, containing $x_i, x_j$ respectively. Thus, the MTF is given by
\begin{equation}
    M = 
    \begin{bmatrix}
    v_{11|x_1\in q_1, x_1 \in q_1} & \dots & v_{1N|x_1\in q_1, x_N \in q_N} \\
    \vdots & \ddots & \vdots \\
    v_{N1|x_N\in q_N, x_1 \in q_1} & \dots & v_{NN|x_N\in q_N, x_N \in q_N} \\
    \end{bmatrix} \\
\end{equation}

Where each value $v_{ij}$ represents the one-step transition probability from $q_i \rightarrow q_j$ in the time series.

The resulting figure visualizes this matrix as a heatmap 
(see Figure \ref{fig:markov}). These fields have been previously used to model temporal dependencies within electroencephalogram (EEG) data
\citep{shankar2022discrimination}.

\subsubsection{Continuous Wavelet Transforms}

The Continuous Wavelet Transform (CWT) allows us to visualize how the frequency components of a time series change over time \citep{grossmann1984decomposition}.

Formally, the Continuous Wavelet Transform (CWT) of a signal \( s(t) \) is defined as:
\begin{equation}
\text{CWT}(s(t))(a, b)=\frac{1}{\sqrt{a}} \int_{-\infty}^{\infty} s(t) \psi^*\left(\frac{t-b}{a}\right) d t
\end{equation}

where \( \text{CWT}(s(t))(a, b) \) represents the CWT of the signal \( s(t) \) with respect to the parameters \( a \) and \( b \),  where \( a \) is the scaling parameter controlling the width of the wavelet function, \( b \) is the translation parameter shifting the wavelet function along the time axis, and \( \psi^*(t) \) is the complex conjugate of the mother wavelet function \( \psi(t) \) used for the transformation.

The resulting plot represents time on the horizontal axis and frequency on the vertical axis. The intensity of colors or shading in the representation indicates the magnitude or power of the different frequencies at each time point. Brighter or more intense colors represent higher power, indicating the presence of a stronger frequency component (See Figure \ref{fig:cwt}). Prior work has found success in representing EEG data with CWTs \citep{khademi2022transfer, narin2022detection}.

\subsubsection{Spectograms}

We generate spectrograms from Short-Time Fourier Transforms (STFT). Like CWT, STFT is a time-frequency analysis technique \citep{allen1977short}. In contrast to CWT, STFT uses fixed windows, and offers either good frequency or time resolution, while CWT uses variable-width wavelets, providing better time-frequency localization and adaptability to non-stationary signals.

Formally, the STFT of a signal \( s(t) \) is defined as:
\begin{equation}
\text{STFT}\{s(t)\}(\tau, \omega) =\int_{-\infty}^{\infty} s(t) w(t-\tau) e^{-i \omega t} d t
\end{equation}
where $w(\tau)$ is the window function and $\omega$ is the frequency. This formula essentially applies a Fourier transform to "windows" of the signal \( x(t) \), where each window is obtained by multiplying \( x(t) \) by the window function \( w(t) \). The window function is usually chosen to be zero outside a certain interval so that the integral is over a finite range.

A spectrogram is an intensity plot of an STFT (see Figure \ref{fig:spectrogram}). Spectrograms are commonly used to represent EEG data \citep{ruffini_deep_2019, kyathanahally_deep_2018}.

\section{Experiments}
\label{sec:experiments}

We evaluated our six representations with  eleven well-known deep learning architectures. Here, we detail the architectures and experimental procedure used. 

\subsection{Models and Finetuning}

\begin{wraptable}{R}{6.5cm}
    \centering
    \begin{tabular}{l|c}
        \toprule
        Model & Parameters\\
        \midrule
        VGG16 \citep{simonyan2014very} & 138M  \\
        EfficientNetB7 \citep{tan2019efficientnet} &  66M  \\
        ResNet152 \citep{he2016deep} & 60.3M  \\
        InceptionResNetV2 \citep{szegedy2017inception} & 56M   \\
        ResNet50 \citep{he2016deep} & 25.6M  \\
        InceptionV3 \citep{szegedy2016rethinking} & 23.9M \\
        Xception \citep{chollet2017xception} & 22.9M  \\
        DenseNet201 \citep{huang2017densely} & 20.2M \\
        EfficientNetB0 \citep{tan2019efficientnet} & 5.3M  \\
        MobileNet \citep{howard2017mobilenets} & 4.2M \\
        MobileNetV2 \citep{howard2017mobilenets} & 3.5M  \\
        \bottomrule
    \end{tabular}
    \caption{Summary of CNN models used.}
    \label{tab:models}
\end{wraptable}

We utilized a selection of CNNs pretrained on the ImageNet dataset, accessible through the Keras Applications library, as the basis for our models. ImageNet is a large-scale dataset of annotated images that is widely used in computer vision research and machine learning \cite{deng_imagenet_2009}. We decided to fine-tune 11 models according to their parameter count ($P$): four large models (with $P$ >= 56M), four medium-sized models (with 20M < $P$ < 56M), and three relatively small models ($P$ < 20M). For an overview of the models we fine-tuned see table \ref{tab:models}. We trained the models with binary cross-entropy loss to predict the presence of a specific artifact. The labels are encoded as one-hot vectors. Details of the training process can be found in \ref{sec:appen_training_details}.


\section{Results}
\label{sec:results}

\begin{figure}[ht]
  \centering
  \begin{minipage}[b]{\textwidth}
    \centering
    \includegraphics[width=\textwidth]{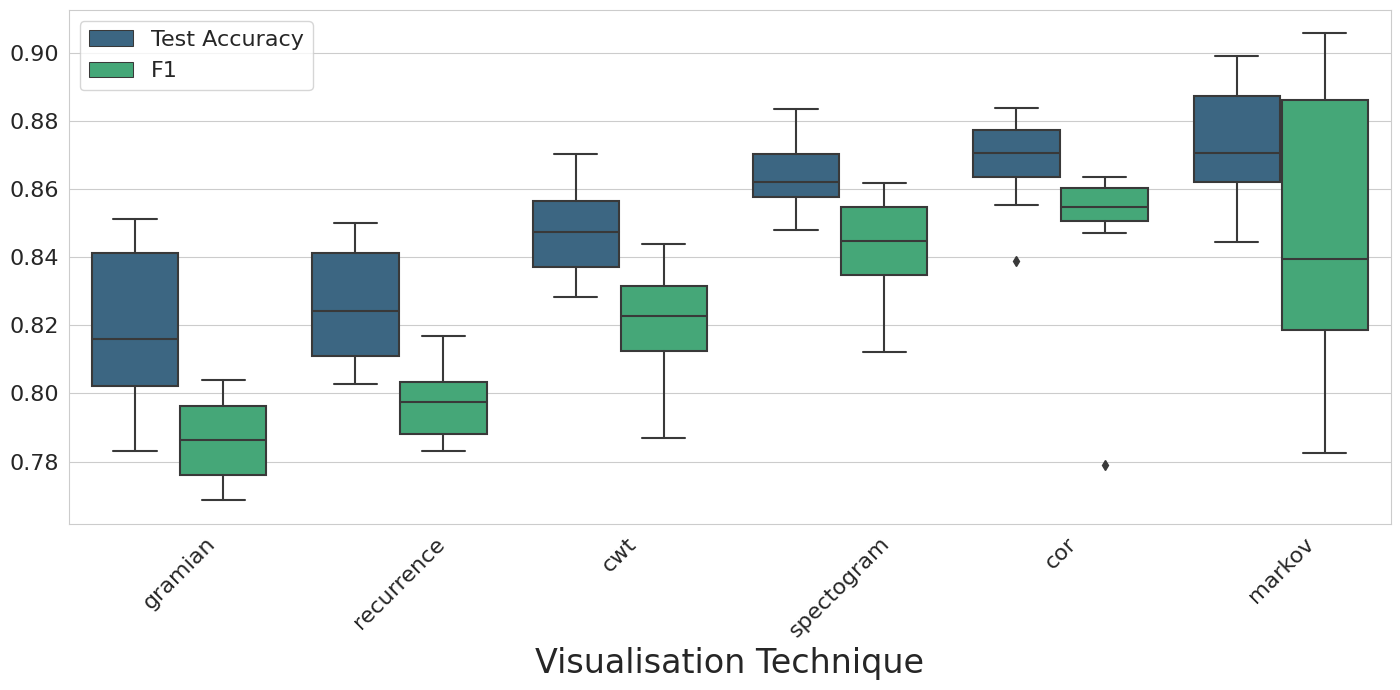}
    \caption{Accuracy and F1 scores on the validation set for each visualization technique, averaged across all models.}
    \label{fig:results}
  \end{minipage}

\end{figure}

The highest performance was achieved with markov transition fields (MTF), closely followed by correlation matrices and spectrograms. 

Xception and EfficientNetB0 emerged as frontrunners in this evaluation. Notably, Xception demonstrated an F1-score of 90.5\% coupled with 89\% accuracy. EfficientNet B0 showed comparable performance achieving an F1-score of 89.4\% and a test accuracy of 89.9\% (see Figure \ref{fig:results}).

F1-scores varied widely across models when MTFs were employed. The superior model exhibited an F1-score that was 12\% higher than the worst-performing model on MTFs. This discrepancy is noteworthy as it is twice the average difference observed between the best and worst-performing models in each testing condition. In comparison to F1, accuracy varied much less with the best model achieving at most 6\% higher scores than others for a given visualization technique. 

For both metrics, the majority of performance differences between visualization techniques were statistically significant at a 5\% significance level, as determined by a paired t-test. Details can be found in \ref{sec:appen_full_results}.

\section{Discussion}
\label{sec:discussion}
Our representations are a classic example of the bias-variance trade-off \citep{neal2019bias}. Some of our representations, such as that of correlation matrices and recurrence plots, contain high bias, and correspondingly, models on these representations fit faster, with a lesser risk of overfitting. Meanwhile, more expressive representations, such as those derived from Fourier transforms, are invertible, and so do not remove any semantic information. However, these representations faithfully represent the noise within the dataset, making downstream models susceptible to overfitting.

It is telling that our top representations lie within the happy medium; in particular, Markov transition matrices are a fairly information-rich representation that, while not fully invertible, was also quite resistant to overfitting.

\subsection{Practitioner's Recommendations}
Within the representations we consider, spectrograms and continuous wavelet transform represent high variance data, while correlation matrices and recurrence plots represent high bias high bias representations. 

In truth, the ideal representation for any given use case will depend on the signal-to-noise ratio of the data. However, within the regimes of low, moderate, and high bias representations, we recommend spectrograms, Markov transition matricies, and correlation matricies, respectively. In line with previous work \citep{Bahador2020Correlation, jalayer2021fault} on these representations, they appear to outperform similar-complexity representations by virtue of capturing and highlighting more salient features within underlying data.

\subsection{Limitations}
This work does not claim to have done a comprehensive assessment of data representation methods; instead, we make strides towards a more nuanced understanding of various image representations of time series data.

As such, it is important to highlight several limitations of our work:
\begin{itemize}

    \item We explored 6 methods for representing sequential data as images, choosing those which we found to be the most popular in prior work. However, there are many such representations which we did not explore, such as mel-spectrograms and cochleagrams, which may be promising.
    \item Similarly, we attempted to make our analysis data representations independent of downstream model choice by profiling performance across 11 well-known CNN model architectures, ranging from 3.5 million to 138.4 million parameters. However, we don't expect our findings to generalize to models outside of CNNs, and similarly, to models much smaller and larger than our explorations.
    \item Finally, while we believe that our choice of a toy problem is useful insofar that it is broadly representative of in-the-wild time series data, we recognize that generalization across domains is a challenging task, and that our findings are unlikely to generalize to data which looks diametrically different from EEG data.
\end{itemize}

\subsection{Future Work}
We believe this work acts as a starting point towards a nuanced understanding of data representation strategies for use in deep learning. Promising directions for future work may include:
\begin{itemize}
    \item Expanding the set of representations considered, and the types and sizes of models considered.
    \item Further comparative analyses on non-image-based representations of time series data, or data representations deriving from non-sequential data.
    \item Considering other toy problems which differ dramatically from EEG data, i.e. time series data without periodic behavior, or with information concentrated within a small band of frequencies.
\end{itemize}

\section{Conclusion}
\label{sec:conclusion}
In this work, we evaluate six image representations of sequential data within a toy problem, artifact detection within EEGs. Three representations, namely Markov Transition Fields, Correlation Matrices, and Spectrograms, have consistently demonstrated superior performance. Given the utility of such representations in improving the performance of downstream machine learning models, we are optimistic that further research in this direction will be useful to the field writ large.

\section{Acknowledgements/Notes}

All authors declare that they have no conflicts of interest.

\pagebreak
\bibliographystyle{elsarticle-num} 
\bibliography{elicit-results}

\begin{thebibliography}{10}
\expandafter\ifx\csname url\endcsname\relax
  \def\url#1{\texttt{#1}}\fi
\expandafter\ifx\csname urlprefix\endcsname\relax\def\urlprefix{URL }\fi
\expandafter\ifx\csname href\endcsname\relax
  \def\href#1#2{#2} \def\path#1{#1}\fi

\bibitem{shorten2019survey}
C.~Shorten, T.~M. Khoshgoftaar, A survey on image data augmentation for deep
  learning, Journal of big data 6~(1) (2019) 1--48.

\bibitem{he2012guided}
K.~He, J.~Sun, X.~Tang, Guided image filtering, IEEE transactions on pattern
  analysis and machine intelligence 35~(6) (2012) 1397--1409.

\bibitem{davis1980comparison}
S.~Davis, P.~Mermelstein, Comparison of parametric representations for
  monosyllabic word recognition in continuously spoken sentences, IEEE
  transactions on acoustics, speech, and signal processing 28~(4) (1980)
  357--366.

\bibitem{grover2016node2vec}
A.~Grover, J.~Leskovec, node2vec: Scalable feature learning for networks, in:
  Proceedings of the 22nd ACM SIGKDD international conference on Knowledge
  discovery and data mining, 2016, pp. 855--864.

\bibitem{mikolov2013efficient}
T.~Mikolov, K.~Chen, G.~Corrado, J.~Dean, Efficient estimation of word
  representations in vector space, arXiv preprint arXiv:1301.3781 (2013).

\bibitem{arias2021multi}
T.~Arias-Vergara, P.~Klumpp, J.~C. Vasquez-Correa, E.~N{\"o}th, J.~R.
  Orozco-Arroyave, M.~Schuster, Multi-channel spectrograms for speech
  processing applications using deep learning methods, Pattern Analysis and
  Applications 24 (2021) 423--431.

\bibitem{yang2019multivariate}
C.-L. Yang, C.-Y. Yang, Z.-X. Chen, N.-W. Lo, Multivariate time series data
  transformation for convolutional neural network, in: 2019 IEEE/SICE
  International Symposium on System Integration (SII), IEEE, 2019, pp.
  188--192.

\bibitem{Bahador2020Correlation}
N.~Bahador, K.~Erikson, J.~Laurila, J.~Koskenkari, T.~Ala-Kokko,
  J.~Kortelainen, A correlation-driven mapping for deep learning application in
  detecting artifacts within the eeg, Journal of Neural Engineering 17~(5)
  (2020) 056018.

\bibitem{michel_eeg_2019}
C.~M. Michel, D.~Brunet,
  \href{https://www.frontiersin.org/articles/10.3389/fneur.2019.00325}{{EEG}
  source imaging: A practical review of the analysis steps} 10.
\newline\urlprefix\url{https://www.frontiersin.org/articles/10.3389/fneur.2019.00325}

\bibitem{jiang_removal_2019}
X.~Jiang, G.-B. Bian, Z.~Tian,
  \href{https://www.ncbi.nlm.nih.gov/pmc/articles/PMC6427454/}{Removal of
  artifacts from {EEG} signals: A review} 19~(5)  987.
\newblock \href {https://doi.org/10.3390/s19050987}
  {\path{doi:10.3390/s19050987}}.
\newline\urlprefix\url{https://www.ncbi.nlm.nih.gov/pmc/articles/PMC6427454/}

\bibitem{tiwary2021time}
H.~Tiwary, A.~Bhavsar, Time-frequency representations for eeg artifact
  classification with cnns, in: 2021 IEEE Applied Imagery Pattern Recognition
  Workshop (AIPR), IEEE, 2021, pp. 1--8.

\bibitem{fu2011review}
T.-c. Fu, A review on time series data mining, Engineering Applications of
  Artificial Intelligence 24~(1) (2011) 164--181.

\bibitem{sadiq2022exploiting}
M.~T. Sadiq, M.~Z. Aziz, A.~Almogren, A.~Yousaf, S.~Siuly, A.~U. Rehman,
  Exploiting pretrained cnn models for the development of an eeg-based robust
  bci framework, Computers in Biology and Medicine 143 (2022) 105242.

\bibitem{xu2019deep}
G.~Xu, X.~Shen, S.~Chen, Y.~Zong, C.~Zhang, H.~Yue, M.~Liu, F.~Chen, W.~Che, A
  deep transfer convolutional neural network framework for eeg signal
  classification, IEEE Access 7 (2019) 112767--112776.

\bibitem{bahador_automatic_2020}
N.~Bahador, K.~Erikson, J.~Laurila, J.~Koskenkari, T.~Ala-Kokko,
  J.~Kortelainen,
  \href{https://ieeexplore.ieee.org/document/9175711/}{Automatic detection of
  artifacts in {EEG} by combining deep learning and histogram contour
  processing}  138--141Conference Name: 2020 42nd Annual International
  Conference of the {IEEE} Engineering in Medicine and Biology Society ({EMBC})
  in conjunction with the 43rd Annual Conference of the Canadian Medical and
  Biological Engineering Society {ISBN}: 9781728119908 Place: Montreal, {QC},
  Canada Publisher: {IEEE}.
\newblock \href {https://doi.org/10.1109/EMBC44109.2020.9175711}
  {\path{doi:10.1109/EMBC44109.2020.9175711}}.
\newline\urlprefix\url{https://ieeexplore.ieee.org/document/9175711/}

\bibitem{hamid_temple_2020}
A.~Hamid, K.~Gagliano, S.~Rahman, N.~Tulin, V.~Tchiong, I.~Obeid, J.~Picone,
  The temple university artifact corpus: An annotated corpus of {EEG}
  artifacts, in: 2020 {IEEE} Signal Processing in Medicine and Biology
  Symposium ({SPMB}), pp. 1--4, {ISSN}: 2473-716X.
\newblock \href {https://doi.org/10.1109/SPMB50085.2020.9353647}
  {\path{doi:10.1109/SPMB50085.2020.9353647}}.

\bibitem{peh_transformer_2022}
W.~Y. Peh, Y.~Yao, J.~Dauwels,
  \href{http://arxiv.org/abs/2208.02405}{Transformer convolutional neural
  networks for automated artifact detection in scalp {EEG}}.
\newblock \href {http://arxiv.org/abs/2208.02405 [eess]}
  {\path{arXiv:2208.02405 [eess]}}.
\newline\urlprefix\url{http://arxiv.org/abs/2208.02405}

\bibitem{ismail2019deep}
H.~Ismail~Fawaz, G.~Forestier, J.~Weber, L.~Idoumghar, P.-A. Muller, Deep
  learning for time series classification: a review, Data mining and knowledge
  discovery 33~(4) (2019) 917--963.

\bibitem{nakano2019effect}
K.~Nakano, B.~Chakraborty, Effect of data representation for time series
  classification—a comparative study and a new proposal, Machine Learning and
  Knowledge Extraction 1~(4) (2019) 1100--1120.

\bibitem{ouyang2008using}
G.~Ouyang, X.~Li, C.~Dang, D.~A. Richards, Using recurrence plot for
  determinism analysis of eeg recordings in genetic absence epilepsy rats,
  Clinical neurophysiology 119~(8) (2008) 1747--1755.

\bibitem{bahari2013eeg}
F.~Bahari, A.~Janghorbani, Eeg-based emotion recognition using recurrence plot
  analysis and k nearest neighbor classifier, in: 2013 20th Iranian Conference
  on Biomedical Engineering (ICBME), IEEE, 2013, pp. 228--233.

\bibitem{wang2015encoding}
Z.~Wang, T.~Oates, et~al., Encoding time series as images for visual inspection
  and classification using tiled convolutional neural networks, in: Workshops
  at the twenty-ninth AAAI conference on artificial intelligence, Vol.~1, AAAI
  Menlo Park, CA, USA, 2015.

\bibitem{islam2019densenet}
M.~M. Islam, M.~M.~H. Shuvo, Densenet based speech imagery eeg signal
  classification using gramian angular field, in: 2019 5th International
  Conference on Advances in Electrical Engineering (ICAEE), IEEE, 2019, pp.
  149--154.

\bibitem{thanaraj2020implementation}
K.~P. Thanaraj, B.~Parvathavarthini, U.~J. Tanik, V.~Rajinikanth, S.~Kadry,
  K.~Kamalanand, Implementation of deep neural networks to classify eeg signals
  using gramian angular summation field for epilepsy diagnosis, arXiv preprint
  arXiv:2003.04534 (2020).

\bibitem{wang2015spatially}
Z.~Wang, T.~Oates, Spatially encoding temporal correlations to classify
  temporal data using convolutional neural networks, arXiv preprint
  arXiv:1509.07481 (2015).

\bibitem{shankar2022discrimination}
A.~Shankar, S.~Dandapat, S.~Barma, Discrimination of types of seizure using
  brain rhythms based on markov transition field and deep learning, IEEE Open
  Journal of Instrumentation and Measurement 1 (2022) 1--8.

\bibitem{grossmann1984decomposition}
A.~Grossmann, J.~Morlet, Decomposition of hardy functions into square
  integrable wavelets of constant shape, SIAM journal on mathematical analysis
  15~(4) (1984) 723--736.

\bibitem{khademi2022transfer}
Z.~Khademi, F.~Ebrahimi, H.~M. Kordy, A transfer learning-based cnn and lstm
  hybrid deep learning model to classify motor imagery eeg signals, Computers
  in biology and medicine 143 (2022) 105288.

\bibitem{narin2022detection}
A.~Narin, Detection of focal and non-focal epileptic seizure using continuous
  wavelet transform-based scalogram images and pre-trained deep neural
  networks, Irbm 43~(1) (2022) 22--31.

\bibitem{allen1977short}
J.~Allen, Short term spectral analysis, synthesis, and modification by discrete
  fourier transform, IEEE Transactions on Acoustics, Speech, and Signal
  Processing 25~(3) (1977) 235--238.

\bibitem{ruffini_deep_2019}
G.~Ruffini, D.~Ibañez, M.~Castellano, L.~Dubreuil-Vall, A.~Soria-Frisch,
  R.~Postuma, J.-F. Gagnon, J.~Montplaisir,
  \href{https://www.frontiersin.org/articles/10.3389/fneur.2019.00806}{Deep
  learning with {EEG} spectrograms in rapid eye movement behavior disorder} 10.
\newline\urlprefix\url{https://www.frontiersin.org/articles/10.3389/fneur.2019.00806}

\bibitem{kyathanahally_deep_2018}
S.~P. Kyathanahally, A.~Döring, R.~Kreis,
  \href{https://onlinelibrary.wiley.com/doi/abs/10.1002/mrm.27096}{Deep
  learning approaches for detection and removal of ghosting artifacts in {MR}
  spectroscopy} 80~(3)  851--863, \_eprint:
  https://onlinelibrary.wiley.com/doi/pdf/10.1002/mrm.27096.
\newblock \href {https://doi.org/10.1002/mrm.27096}
  {\path{doi:10.1002/mrm.27096}}.
\newline\urlprefix\url{https://onlinelibrary.wiley.com/doi/abs/10.1002/mrm.27096}

\bibitem{simonyan2014very}
K.~Simonyan, A.~Zisserman, Very deep convolutional networks for large-scale
  image recognition, arXiv preprint arXiv:1409.1556 (2014).

\bibitem{tan2019efficientnet}
M.~Tan, Q.~Le, Efficientnet: Rethinking model scaling for convolutional neural
  networks, in: International conference on machine learning, PMLR, 2019, pp.
  6105--6114.

\bibitem{he2016deep}
K.~He, X.~Zhang, S.~Ren, J.~Sun, Deep residual learning for image recognition,
  in: Proceedings of the IEEE conference on computer vision and pattern
  recognition, 2016, pp. 770--778.

\bibitem{szegedy2017inception}
C.~Szegedy, S.~Ioffe, V.~Vanhoucke, A.~Alemi, Inception-v4, inception-resnet
  and the impact of residual connections on learning, in: Proceedings of the
  AAAI conference on artificial intelligence, Vol.~31, 2017.

\bibitem{szegedy2016rethinking}
C.~Szegedy, V.~Vanhoucke, S.~Ioffe, J.~Shlens, Z.~Wojna, Rethinking the
  inception architecture for computer vision, in: Proceedings of the IEEE
  conference on computer vision and pattern recognition, 2016, pp. 2818--2826.

\bibitem{chollet2017xception}
F.~Chollet, Xception: Deep learning with depthwise separable convolutions, in:
  Proceedings of the IEEE conference on computer vision and pattern
  recognition, 2017, pp. 1251--1258.

\bibitem{huang2017densely}
G.~Huang, Z.~Liu, L.~Van Der~Maaten, K.~Q. Weinberger, Densely connected
  convolutional networks, in: Proceedings of the IEEE conference on computer
  vision and pattern recognition, 2017, pp. 4700--4708.

\bibitem{howard2017mobilenets}
A.~G. Howard, M.~Zhu, B.~Chen, D.~Kalenichenko, W.~Wang, T.~Weyand,
  M.~Andreetto, H.~Adam, Mobilenets: Efficient convolutional neural networks
  for mobile vision applications, arXiv preprint arXiv:1704.04861 (2017).

\bibitem{deng_imagenet_2009}
J.~Deng, W.~Dong, R.~Socher, L.-J. Li, K.~Li, L.~Fei-Fei, {ImageNet}: A
  large-scale hierarchical image database, in: 2009 {IEEE} Conference on
  Computer Vision and Pattern Recognition, pp. 248--255, {ISSN}: 1063-6919.
\newblock \href {https://doi.org/10.1109/CVPR.2009.5206848}
  {\path{doi:10.1109/CVPR.2009.5206848}}.

\bibitem{neal2019bias}
B.~Neal, On the bias-variance tradeoff: Textbooks need an update, arXiv
  preprint arXiv:1912.08286 (2019).

\bibitem{jalayer2021fault}
M.~Jalayer, C.~Orsenigo, C.~Vercellis, Fault detection and diagnosis for
  rotating machinery: A model based on convolutional lstm, fast fourier and
  continuous wavelet transforms, Computers in Industry 125 (2021) 103378.

\bibitem{kingma2014adam}
D.~P. Kingma, J.~Ba, Adam: A method for stochastic optimization, arXiv preprint
  arXiv:1412.6980 (2014).

\end{thebibliography}


\pagebreak
\appendix

\section{Full Results}
\label{sec:appen_full_results}

\begin{longtable}{llrrrr}

\toprule
visualization & Model &  Accuracy &  Precision &   Recall &       F1 \\
\midrule
\endhead
\multirow{1}{*}{cor} & densenet201 & 0.855 & 0.850 & 0.844 & 0.847 \\
\multirow{58}{*}{} & \textbf{efficientnetb0*} & 0.875 & \textbf{0.863} & 0.860 & 0.862 \\
 & \textbf{efficientnetb7*} & 0.875 & 0.853 & \textbf{0.868} & 0.861 \\
 & \textbf{inceptionresnetv2*} & \textbf{0.884} & 0.861 & 0.866 & \textbf{0.864} \\
 & inceptionv3 & 0.879 & 0.842 & 0.866 & 0.854 \\
 & mobilenet & 0.871 & 0.846 & 0.864 & 0.855 \\
 & mobilenetv2 & 0.868 & 0.855 & 0.846 & 0.850 \\
 & resnet152 & 0.860 & 0.847 & 0.855 & 0.851 \\
 & resnet50 & 0.867 & 0.846 & 0.865 & 0.855 \\
 & vgg16 & 0.839 & 0.839 & 0.727 & 0.779 \\
 & xception & 0.880 & 0.853 & 0.867 & 0.860 \\
  \hline
\multirow{1}{*}{cwt} & \textbf{densenet201*} & \textbf{0.870} & \textbf{0.861} & 0.828 & \textbf{0.844} \\
 & \textbf{efficientnetb0*} & 0.862 & 0.855 & \textbf{0.832} & 0.843 \\
 & efficientnetb7 & 0.849 & 0.834 & 0.823 & 0.828 \\
 & inceptionresnetv2 & 0.832 & 0.791 & 0.830 & 0.810 \\
 & inceptionv3 & 0.842 & 0.835 & 0.800 & 0.817 \\
 & mobilenet & 0.843 & 0.830 & 0.816 & 0.823 \\
 & mobilenetv2 & 0.865 & 0.854 & 0.816 & 0.835 \\
 & resnet152 & 0.828 & 0.818 & 0.801 & 0.809 \\
 & resnet50 & 0.850 & 0.839 & 0.809 & 0.824 \\
 & vgg16 & 0.847 & 0.847 & 0.734 & 0.787 \\
 & xception & 0.828 & 0.804 & 0.826 & 0.814 \\
  \hline
\multirow{1}{*}{gramian} & densenet201 & 0.847 & 0.832 & 0.765 & 0.797 \\
 & efficientnetb0 & 0.807 & 0.806 & 0.782 & 0.794 \\
 & \textbf{efficientnetb7*} & 0.839 & 0.836 & \textbf{0.794} & \textbf{0.815} \\
 & \textbf{inceptionresnetv2*} & \textbf{0.851} & 0.839 & 0.771 & 0.804 \\
 & inceptionv3 & 0.804 & 0.800 & 0.752 & 0.775 \\
 & mobilenet & 0.783 & 0.773 & 0.765 & 0.769 \\
 & mobilenetv2 & 0.825 & 0.819 & 0.779 & 0.799 \\
 & resnet152 & 0.831 & 0.822 & 0.762 & 0.791 \\
 & resnet50 & 0.802 & 0.794 & 0.763 & 0.778 \\
 & \textbf{vgg16*} & 0.845 & \textbf{0.842} & 0.730 & 0.782 \\
 & xception & 0.784 & 0.762 & 0.778 & 0.770 \\
  \hline
\multirow{1}{*}{markov} & densenet201 & 0.857 & 0.854 & 0.779 & 0.814 \\
 & \textbf{efficientnetb0**} & \textbf{0.899} & \textbf{0.903} & 0.886 & 0.894 \\
 & efficientnetb7 & 0.898 & 0.900 & 0.890 & 0.895 \\
 & inceptionresnetv2 & 0.881 & 0.880 & 0.839 & 0.859 \\
 & inceptionv3 & 0.861 & 0.858 & 0.780 & 0.817 \\
 & mobilenet & 0.884 & 0.887 & 0.869 & 0.878 \\
 & mobilenetv2 & 0.871 & 0.872 & 0.809 & 0.839 \\
 & resnet152 & 0.863 & 0.865 & 0.780 & 0.820 \\
 & resnet50 & 0.864 & 0.866 & 0.813 & 0.839 \\
 & vgg16 & 0.844 & 0.844 & 0.729 & 0.782 \\
 & \textbf{xception**} & 0.891 & 0.896 & \textbf{0.916} & \textbf{0.906} \\
 \hline
\multirow{1}{*}{recurrence} & densenet201 & 0.850 & 0.839 & 0.786 & 0.811 \\
 & efficientnetb0 & 0.817 & 0.806 & 0.799 & 0.802 \\
 & \textbf{efficientnetb7*} & \textbf{0.855} & \textbf{0.854} & \textbf{0.814} & \textbf{0.834} \\
 & inceptionresnetv2 & 0.835 & 0.829 & 0.780 & 0.804 \\
 & inceptionv3 & 0.826 & 0.819 & 0.777 & 0.798 \\
 & mobilenet & 0.822 & 0.799 & 0.795 & 0.797 \\
 & mobilenetv2 & 0.846 & 0.840 & 0.795 & 0.817 \\
 & resnet152 & 0.803 & 0.802 & 0.771 & 0.787 \\
 & resnet50 & 0.804 & 0.789 & 0.791 & 0.790 \\
 & vgg16 & 0.843 & 0.843 & 0.731 & 0.783 \\
 & xception & 0.809 & 0.770 & 0.805 & 0.787 \\
 \hline
\multirow{1}{*}{spectogram} & \textbf{densenet201*} & \textbf{0.883} & 0.812 & 0.854 & 0.833 \\
 & \textbf{efficientnetb0*} & 0.870 & 0.869 & 0.855 & \textbf{0.862} \\
 & \textbf{efficientnetb7*} & 0.871 & 0.859 & \textbf{0.860} & 0.860 \\
 & inceptionresnetv2 & 0.860 & 0.819 & 0.855 & 0.837 \\
 & inceptionv3 & 0.848 & 0.845 & 0.835 & 0.840 \\
 & mobilenet & 0.862 & 0.852 & 0.852 & 0.852 \\
 & \textbf{mobilenetv2*} & 0.875 & \textbf{0.877} & 0.839 & 0.858 \\
 & resnet152 & 0.862 & 0.853 & 0.837 & 0.845 \\
 & vgg16 & 0.849 & 0.822 & 0.843 & 0.833 \\
 & xception & 0.855 & 0.856 & 0.772 & 0.812 \\
\bottomrule
\caption{The performance of each combination of model and data representation. Models that perform best on a single metric for each representation are highlighted with \textbf{*}. Models that perform best on a single metric overall are highlighted with \textbf{**}.} \\
\end{longtable}

\subsection{Comparing Performance across Models}

\begin{figure}[ht]
  \centering
  \begin{minipage}[b]{\textwidth}
    \centering
    \includegraphics[width=\textwidth]{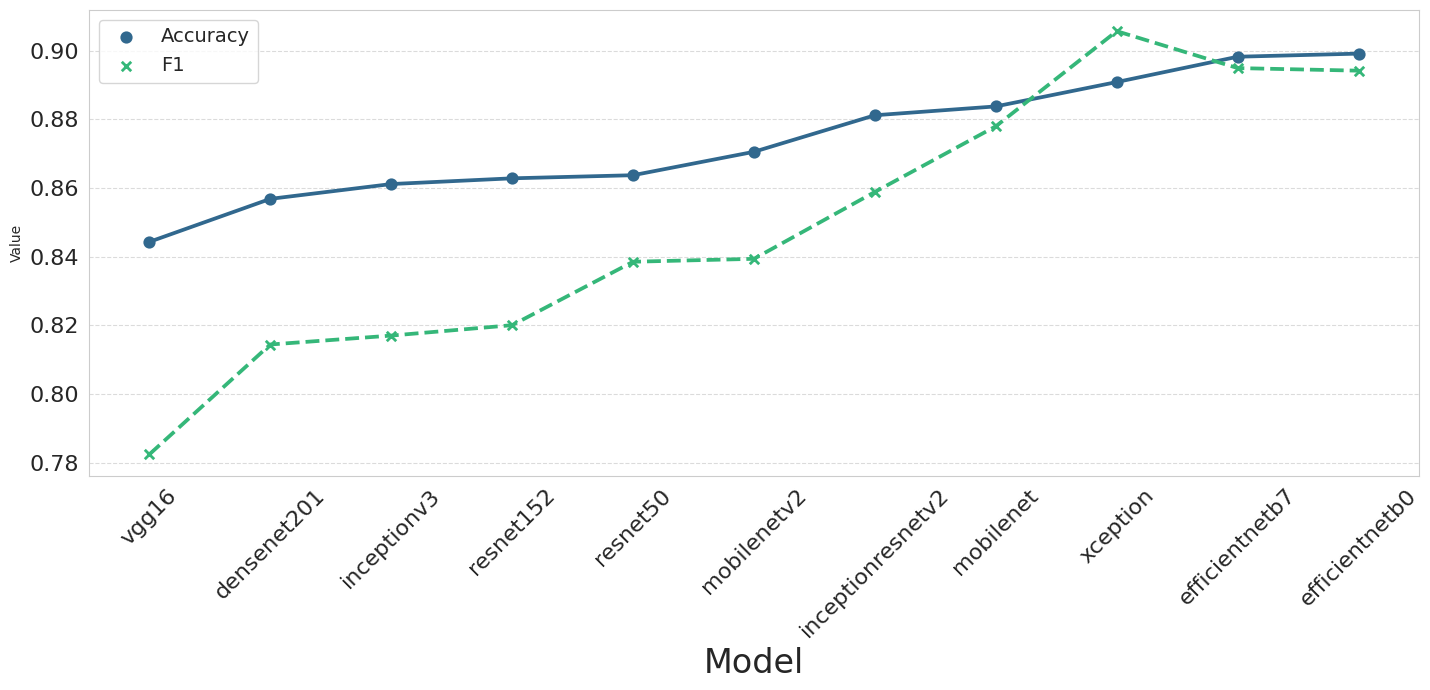}
    \caption{Comparing the performance of different architectures on MTFs.}
    \label{fig:appen_within_representations}
  \end{minipage}

\end{figure}

Here, we examine the effect of different architectures on performance, focusing on Markov Transition Fields (see Figure \ref{fig:appen_within_representations}). For this representation, Xception and EfficientNet models performed best, both in terms of accuracy and F1 score. 

\begin{figure}[ht]

\centering
\includegraphics[width=\textwidth]{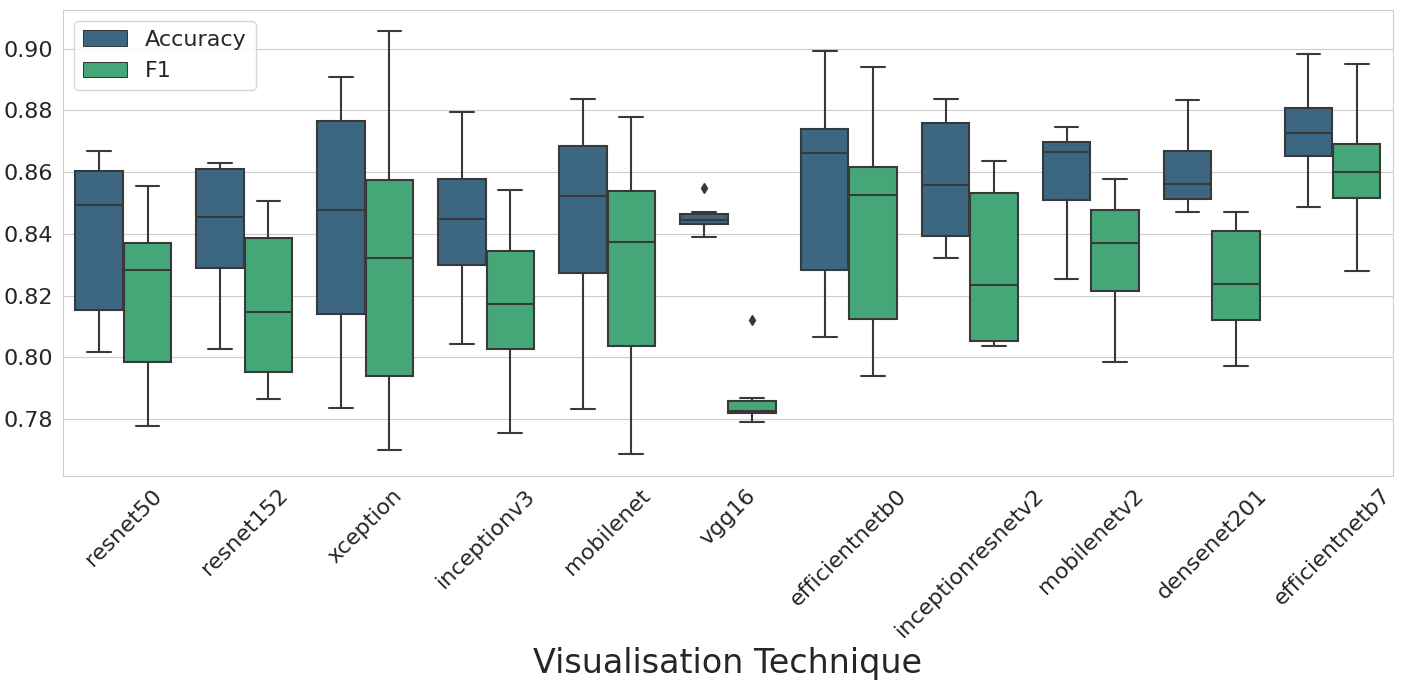}
\caption{Comparing F1 and Accuracy across models, averaged over visualization techniques.}
\label{fig:appen_intermodel}
\end{figure}

Next, we look at the performance differences between models when averaged over all data representations. Some models' performance was much more sensitive to data representations than others'; VGG16 was remarkably consistent in its performance, while Xception was $10\%$ more accurate on some representations than on others (see Figure \ref{fig:appen_intermodel}). 

\subsection{Comparing Performance across Data Representations}

\begin{figure}[ht]
  \centering
  \begin{minipage}[b]{\textwidth}
    \centering
    \includegraphics[width=\textwidth]{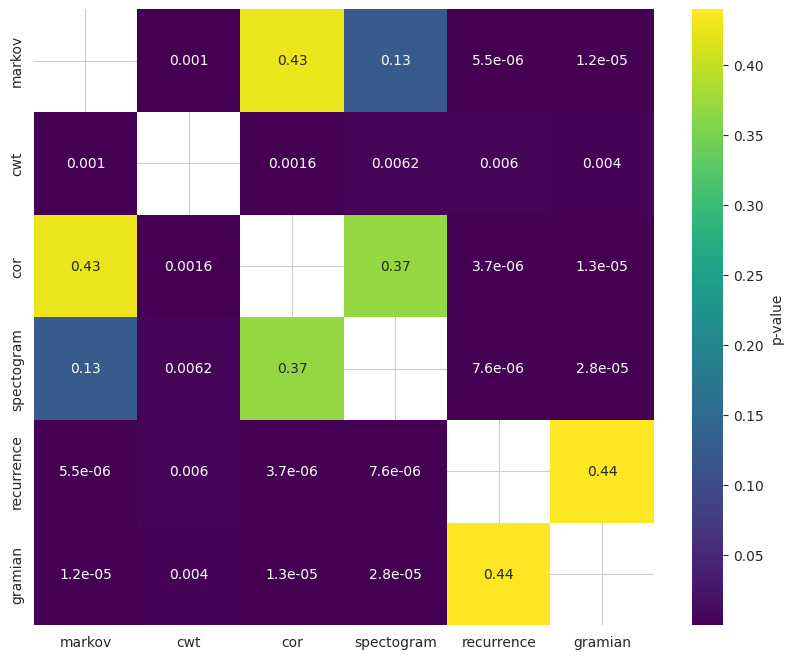}
    \caption{Assessing the statistical significance of average differences in accuracy between visualization techniques. The analysis was performed with a paired t-test.}
    \label{fig:appen_p-value_heatmap}
  \end{minipage}

\end{figure}

The majority of average performance differences in accuracy and F1 scores between visualization techniques were statistically significant at a 5\% significance level, as determined by a paired t-test (see Figure \ref{fig:appen_p-value_heatmap}).  We found four combinations were the difference was not statistically significant: MTFs and spectrograms, MTFs and correlation matrices, correlation matrices and spectrograms and, lastly recurrence plots and GASFs.

\section{Training details}
\label{sec:appen_training_details}
Each pretrained model was modified in two ways: first, we prepend a convolutional layer to convert the image representations into image features, and secondly, we replace the ImageNet classification layer with a global average pooling layer and a 5-node fully-connected layer with sigmoid activation for our classification task. 

Our training process involves a training stage and a fine-tuning stage. In the training stage, we keep the pretrained model backbone frozen, and only train the added layers for $5$ epochs, utilizing the Adam optimizer \citep{kingma2014adam} with a learning rate of $0.001$. In the finetuning stage, we unfreeze all the pretrained model layers except the batch normalization layers, and continue training until convergence with a learning rate of $0.00001$.

\end{document}